\documentclass[prl,twocolumn,showpacs,preprintnumbers,amsmath,amssymb]{revtex4}

\usepackage{graphicx}
\usepackage{dcolumn}
\usepackage{bm}

\newcommand{\xx}{{\bm x}}
\newcommand{\nn}{{\bm n}}

\begin{document}

\preprint{APS/123-QED}

\title{Knots in a Spinor Bose-Einstein Condensate}
\author{Yuki Kawaguchi$^1$}
\author{Muneto Nitta$^2$}
\author{Masahito Ueda$^{1,3}$}
\affiliation{$^1$Department of Physics, Tokyo Institute of Technology,
2-12-1 Ookayama, Meguro-ku, Tokyo 152-8551, Japan \\
$^2$Department of Physics, Keio University, Hiyoshi, Yokohama, Kanagawa 223-8521, Japan\\
$^3$ERATO Macroscopic Quantum Control Project, JST, Tokyo 113-8656, Japan
}
\date{\today}

\begin{abstract}
We show that knots of spin textures can be created in the polar phase of a
 spin-1 Bose-Einstein condensate, and discuss experimental schemes for
 their generation and probe, together with their lifetime.
\end{abstract}


\maketitle


A decade ago, Faddeev and Niemi suggested that knots might exist as stable solitons in a three-dimensional classical field theory,
thus opening up a way to investigate physical properties of knot-like structures~\cite{Faddeev1997}.
They further proposed \cite{Faddeev1999} that their model can be interpreted as the low-energy limit of the Yang-Mills theory,
where knots are suggested as a natural candidate for describing glueballs -- massive particles made of gluons.

In cosmology, topological defects are considered to be important for understanding the large-scale structure of our universe \cite{VS}.
Although recent measurements of the cosmic microwave background (CMB) have shown that topological defects are not the dominant source of CMB anisotropies,
the search for topological objects in the universe, such as cosmic strings, continues to be actively conducted.
Recently, it has been suggested that a cosmic texture, classified by the homotopy group of $\pi_3(S^3)=\mathbb{Z}$ \cite{VS},
generates cold and hot spots of CMB \cite{Cruz2007}.
This texture is a spherical or point-like object that is unstable against shrinkage according to the scaling argument.
In cosmology, the instability is favored
for evading cosmological problems such as a monopole problem.
Three-dimensional skyrmions~\cite{Skyrmion} and Shankar monopoles~\cite{Shankar1977,Volovik1977b} are
topological objects belonging to the same homotopy group.
On the other hand, knots, which belong to a distinct homotopy group, $\pi_3(S^2)=\mathbb{Z}$,
have thus far been ignored by cosmologists; however, they would be a potential candidate for topological solitons in our universe.

Knots are unique topological objects characterized by a linking number
or a Hopf invariant as discussed in a seminal paper on superfluid $^3$He~\cite{Volovik1977b}.
Other familiar topological objects, such as vortices~\cite{deGennes1973,VolovikMineev,Ho1978,Mermin1979}, monopoles~\cite{tHooftPolyakov,VolovikMineev}, 
and skyrmions~\cite{Skyrmion,Shankar1977,Volovik1977b},
are characterized by winding numbers,
which have recently been discussed in relation to spinor Bose-Einstein condensates
(BECs)~\cite{Leonhardt2000,Zhou2001,Makela2003,Barnett2007,Stoof2001,Anglin2003,KhawajaStoof}.
However, little is known about how to create such knots experimentally.
In this Letter, we point out that spinor BECs offer an ideal testing ground for investigating the dynamic creation and destruction of knots.
We also show that knots can be imprinted in an atomic BEC using conventional magnetic field configurations. 

We consider a BEC of spin-1 atoms with mass $M$ that are trapped in an optical potential $V(\xx)$.
The energy functional for a BEC at zero magnetic field is given by
\begin{align}
E = &\int d^3 x \bigg[ \frac{\hbar^2}{2M}\sum_m|{\bm \nabla}\psi_m|^2
 + V\rho + \frac{g_0}{2}\rho^2 + \frac{g_1}{2}\rho^2 |{\bm f}|^2 \bigg],
\end{align}
where $\psi_m(\xx)$ is an order parameter of the BEC in a magnetic sublevel $m=1, 0$ or $-1$ at position $\xx$,
and $\rho(\xx)=\sum_m |\psi_m(\xx)|^2$ and
$\rho(\xx){\bm f}(\xx)=\sum_{mm'}\psi^\ast_m(\xx){\bm F}_{mm'}\psi_{m'}(\xx)$
are the number density and spin density, respectively.
Here ${\bm F}=(F_x, F_y, F_z)$ is a vector of spin-1 matrices.
The strength of the interaction is given by
$g_0=4\pi\hbar^2(2a_2+a_0)/(3M)$ and $g_1=4\pi\hbar^2(a_2-a_0)/(3M)$,
where $a_S$ is the {\it s}-wave scattering length for two colliding atoms
with total spin $S$.
The ground state is polar for $g_1>0$ and ferromagnetic for $g_1<0$~\cite{Ohmi1998, Ho1998}.

The order parameter for the polar phase can be described by the
superfluid phase $\vartheta$ and unit vector field ${\bm n}$ in spin space, whose components are given in terms of $\psi_m$ as
\begin{align}
{\bm \Psi} \equiv \begin{pmatrix} \psi_1 \\ \psi_0 \\ \psi_{-1}\end{pmatrix}
 = \sqrt{\rho}\, e^{i\vartheta} \begin{pmatrix}
\frac{-n_x+in_y}{\sqrt{2}}\\ n_z \\ \frac{n_x+in_y}{\sqrt{2}}\end{pmatrix}.
\label{eq:OP}
\end{align}
This order parameter is invariant under an arbitrary rotation about $\nn=(n_x,n_y,n_z)$, 
i.e., $\exp[-i{\bm n}\cdot {\bm F} \gamma]{\bm \Psi}={\bm \Psi}$,
where $\gamma$ is an arbitrary real number.
It is also invariant under simultaneous transformations $\vartheta \to \vartheta + \pi$ and $\nn \to -\nn$.
The order parameter manifold for the polar phase is therefore given by $M = {(U(1) \times S^2) /\mathbb{Z}_2}$~\cite{Zhou2001},
where $U(1)$ denotes the manifold of the superfluid phase $\vartheta$,
and $S^2$ is a two-dimensional sphere whose point specifies the direction of $\nn$.

Knots are characterized by mappings from a three-dimensional sphere $S^3$ to $S^2$.
The $S^3$ domain is prepared by imposing a boundary condition that ${\bm \Psi}$ takes on the same value 
in every direction at spatial infinity, so that 
the medium is compactified into $S^3$.
Since neither $U(1)$ nor $\mathbb{Z}_2$ symmetry contributes to homotopy groups in spaces higher than one dimension, 
we have $\pi_3(M)\cong \pi_3(S^2)=\mathbb{Z}$.
The associated integer topological charge $Q$ is known as the Hopf charge which is given by
\begin{align}
Q=\frac{1}{4\pi^2} \int d^3x\ \epsilon_{ijk}\mathcal{F}_{ij} \mathcal{A}_k,
\label{eq:H-charge}
\end{align}
where $\mathcal{F}_{ij}=\partial_i\mathcal{A}_j-\partial_j\mathcal{A}_i=\nn\cdot(\partial_i \nn \times \partial_j \nn)$~\cite{Faddeev1997}.
Note that the domain ($\xx$) is three-dimensional, while the target space ($\nn$) is two-dimensional.
Consequently, the preimage of a point on the target $S^2$ constitutes a closed loop in $S^3$,
and the Hopf charge is interpreted as the linking number of these loops:
If the $\nn$ field has Hopf charge $Q$,
two loops corresponding to the preimages of any two distinct points on the target $S^2$ will be linked $Q$ times.

Figure~1~(a) illustrates the $\nn$ field of a knot created in a uniform BEC with the linking number of 1.
Here, the $\nn$ field is expressed as $\nn(\xx)=\exp\left[ -i \alpha(r)\hat{\bm v}(\theta,\phi)\cdot \widetilde{\bm F}\right]\nn_\infty$,
subject to the boundary condition $\nn_\infty=(0,0,1)^T$ at spatial infinity, 
where $\xx=(r\sin\theta\cos\phi, r\sin\theta\sin\phi, r\cos\theta)$
and $\widetilde{\bm F}$ is a vector of spin-1 matrices in the Cartesian representation, i.e.,
$\widetilde{F}_j=-i\epsilon_{jkl}$ $(j,k,l=x,y,z)$.
A given mapping $\hat{\bm v}(\theta,\phi) : S^2 \to S^2$ determines the charge $Q$,
and we choose $\hat{\bm v}=(-\sin\theta\cos\phi,-\sin\theta\sin\phi,\cos\theta)$ in Fig.~1~(a).
The radial profile function $\alpha(r)$ is a monotonically decreasing function of $r$, subject to the boundary conditions
$\alpha(0) = 2\pi$ and $\alpha(\infty) = 0$. 
Here, we consider $\alpha(r)=2\pi[1-\tanh(r/\xi_{\rm knot})]$, where $\xi_{\rm knot}$ is a characteristic size of the knot.
Although there is no singularity in the texture, it is impossible to wind it off to a uniform configuration
because this texture has a nonzero Hopf charge of 1.
The left part of Fig.~1 (b) describes the order parameter for the $m=-1$ component in real space,
where we plot the isopycnic surface of the density and the color on the surface represents the phase.
Here, $\nn$ on the torus is almost perpendicular to $\nn_\infty$ and the phase of $\psi_{-1}$ is given by $\arg(\psi_{-1})=\arctan(n_y/n_x)$.
The core of the knot ($\nn=-\nn_\infty$) is depicted as a white tube in Fig.~1(b).
On the right side of the figure, we show the extracted preimage for $\nn=(0,0,-1)^T$ (white tube) and that for $\nn=(1,0,0)^T$ (black tube).

\begin{figure}
\includegraphics[width=0.87\linewidth]{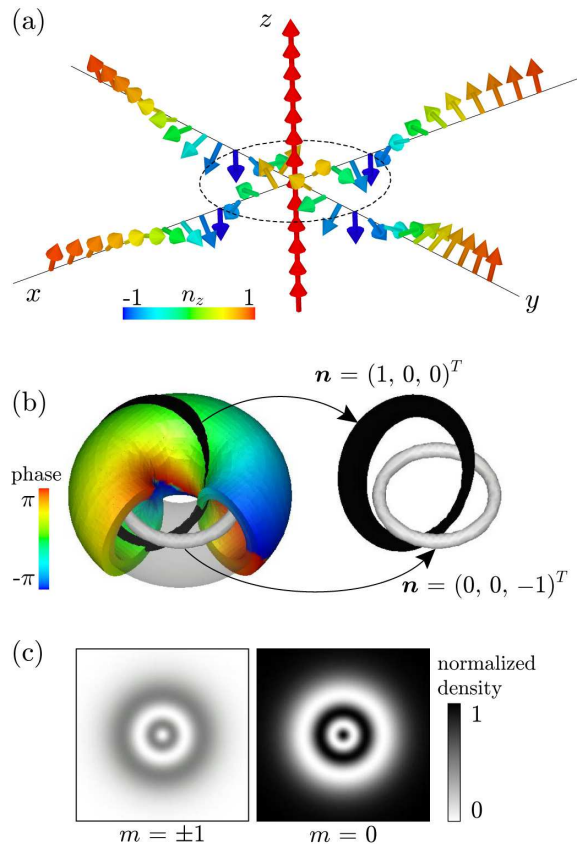}
\caption{(Color)
(a) Three-dimensional configuration of the $\nn$ field of a knot with Hopf charge $Q=1$, expressed as
$\nn = e^{-i\alpha(r)\hat{\bm v}\cdot \widetilde{\bm F}}(0, 0, 1)^T$,
where $\alpha(r)=2\pi[1- \tanh(r/\xi_{\rm knot})]$ and $\hat{\bm v}=(-\sin\theta\cos\phi, -\sin\theta\sin\phi,\cos\theta)$.
Note that $\nn=(n_x, n_y, n_z)^T$ is related to the order parameter ${\bm \Psi}$ through Eq.~\eqref{eq:OP}.
In the figure, only $\nn$ on the $x, y$, and $z$ axes is shown as arrows.
The dashed line traces the point $\nn=(0,0,-1)^T$, which is a circle on the $xy$ plane with the radius $\left[\tanh^{-1}(1/2)\right]\xi_{\rm knot}\simeq 0.55\,\xi_{\rm knot}$.
The color on the arrows represents the value of $n_z$ (see the scale).
(b) The torus shape on the left side of the figure shows the isopycnic surface of $|\psi_{-1}|^2=0.47$,
which is equivalent to the region where $|n_z|\le 0.24$.
The color on the surface shows the phase of $\psi_{-1}$, that is equivalent to $\arctan (n_y/n_x)$.
The white and black tubes are the preimages of $\nn\simeq(0,0,-1)^T~ (n_z<-0.95)$ and $\nn\simeq(1,0,0)^T~(n_x>0.95)$, respectively,
which are reproduced on the right side of the figure.
The two tubes cross once (linking number 1), which is consistent with $Q=1$.
(c) Cross sections of the density of the $m=\pm 1$ (left) and $m=0$ (right) components in the $xy$ plane.
The size of each panel is $4 \xi_{\rm knot} \times 4 \xi_{\rm knot}$.
The torus shape of the $m=\pm 1$ components appears as double rings in the cross section, which can serve as the signature of a knot.
}
\label{fig:1}
\end{figure}

The torus shape of the $m=-1$ component appears as a double-ring pattern in the cross-sectional plane at $z=0$, as shown in Fig.~1~(c), 
where the density distributions of the $m=\pm 1$ (left) and $m=0$ (right) components are shown in the gray scale.
The distributions of $m=\pm1$ components overlap completely; therefore the system remains unmagnetized.
This double-ring pattern can serve as an experimental signature of a knot 
and it should be probed by performing the Stern-Gerlach experiment on the BEC that is sliced at $z=0$.

Next, we show that knots can be created by manipulating an external magnetic field.
In the presence of an external magnetic field,
the time-dependent phase differences between different spin components are induced because of the linear Zeeman effect,
which causes the Larmor precession of $\nn$,
while $\nn$ tends to become parallel to the magnetic field because of the quadratic Zeeman effect.
Suppose that we prepare a BEC in the $m=0$ state, i.e., $\nn=(0,0,1)^T$,
by applying a uniform magnetic field ${\bm B}_0$ in the $z$ direction.
Then, we suddenly turn off ${\bm B}_0$ and switch on a magnetic field,
${\bm B}(\xx) = b(r)\,(\sin\theta\cos(\phi+\phi_0),\sin\theta \sin(\phi+\phi_0),\cos\theta)^T$,
where $b(r)$ is an arbitrary function and $\phi_0$ is a real parameter.
This magnetic field configuration is quadrupolar if $\phi_0=\pi$ and monopolar if $\phi_0=0$.
In what follows, we shall consider the case of $\phi_0=\pi$ and $b(r)=b' r$, unless otherwise specified. 
Because of the linear Zeeman effect, $\nn$ starts rotating around the local magnetic field as
$\nn = \exp[-i (\mu_{\rm B}/2){\bm B}(\xx)\cdot \widetilde{\bm F}\, t/\hbar] (0, 0, 1)^T$,
and therefore the $\nn$ field winds as a function of $t$.

Figure~2 shows the dynamics of the creation and destruction of knots in a spherical trap subject to the quadrupole field.
Figures~2~(a)--(d) show the snapshots of $\psi_{-1}$ (left) and preimages of $\nn=(0,0,-1)^T$ and $\nn=(1,0,0)^T$ (right)
at (a) $t=0.5 T_{\rm L}$, (b) $1.1 T_{\rm L}$, (c) $2.2 T_{\rm L}$ and (d) $3.3 T_{\rm L}$,
where $T_{\rm L}=2h/(\mu_B b' R_{\rm TF})$ is the period of the Larmor precession at the Thomas-Fermi radius $R_{\rm TF}$.
The solid circle indicates the periphery of the BEC on the $xy$ plane.
Knot-like objects enter the BEC from its periphery [Fig.~2~(a)],
and the number of knots increases as the $\nn$ field winds more and more with time [Fig.~2~(b), (c)].
With the passage of further time, however, the knot structure is destroyed, as shown in Fig.~2~(d).
The knots therefore have a finite lifetime, which will be discussed later.
Figures~2~(e)--(h) show cross sections of the density for $m=-1$ (top) and $m=0$ (bottom) components on the $xy$ plane
at (e) $t=0.5 T_{\rm L}$, (f) $1.1 T_{\rm L}$, (g) $2.2 T_{\rm L}$, and (h) $3.3 T_{\rm L}$.
We find that as the $\nn$ field winds with time, the number of rings increases.
This prediction can be tested by the Stern-Gerlach experiment.

\begin{figure}
\includegraphics[width=0.87\linewidth]{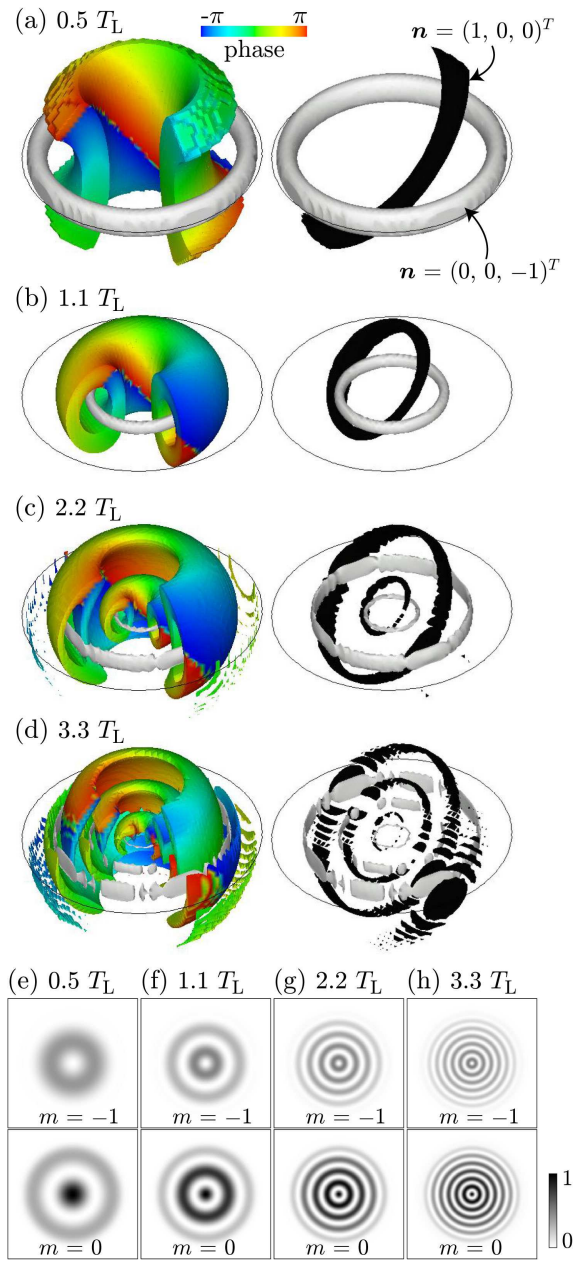}
\caption{(Color) Dynamics of the creation and destruction of knots in a spherical trap under a quadrupolar magnetic field.
(a)--(d): The left sides of the figure show the isopycnic surface of $|\psi_{-1}|^2/\rho=0.4$, and the color on the surface shows the phase of $\psi_{-1}$
at (a) $t= 0.5T_{\rm L}$, (b) $1.1T_{\rm L}$, (b) $2.2T_{\rm L}$, and (d) $3.3T_{\rm L}$, where $T_{\rm L}=2h/(\mu_B b' R_{\rm TF})$.
The white and black tubes on the right are the preimages of $\nn\simeq(0,0,-1)^T~ (n_z<-0.95)$ and $\nn\simeq(1,0,0)^T~ (n_x>0.95)$, respectively,
and the solid circles represent the periphery of the BEC [$\rho/\rho(\xx=0)=0.01$] in the $xy$ plane.
(e)--(h) Cross sections of the density normalized by $\rho(\xx=0)$ for $m=-1$ (top) and $m=0$ (bottom) components in the $xy$ plane
at (e) $t= 0.5T_{\rm L}$, (f) $1.1T_{\rm L}$, (g) $2.2T_{\rm L}$, and (h) $3.3T_{\rm L}$.
}
\end{figure}

The knot soliton in the simple nonlinear $\sigma$ model without a higher derivative term is known to be energetically unstable,
since the energy of the knot is proportional to its size $\xi_{\rm knot}$,
which can be calculated by integrating the kinetic energy $\sim \xi_{\rm knot}^{-2}$ in the
volume of the soliton $\sim \xi_{\rm knot}^3$.
Knots will therefore shrink and finally disappear.
In the case of an atomic gaseous BEC, however, 
the total energy of the system has to be conserved; therefore,
the above-mentioned energetics argument does not imply the instability of knots in trapped systems.

The dominant mechanism for destroying the knot in the spinor BEC is the spin current caused by the spatial dependence of the $\nn$ field.
The spin current induces local magnetization according to
\begin{align}
 \frac{\partial f_\mu }{\partial t} = - {\bm \nabla}\cdot{\bm j}_{\rm spin}^\mu = \frac{\hbar}{M} \sum_{\nu\lambda}\epsilon_{\mu\nu\lambda} n_\nu \nabla^2 n_\lambda,
\label{eq:dfdt}
\end{align}
thereby destroying the polar state.
The initial polar state will spontaneously develop into a biaxial nematic state,
and eventually result in a fully polarized ferromagnetic domain.
While $\nn$ is well-defined in a biaxial nematic state as one of the symmetry axes,
it is ill-defined in the fully-magnetized region.

Substituting $\nn = e^{-i\alpha(r)\hat{\bm v}\cdot \widetilde{\bm F}}(0, 0, 1)^T$
as used for Fig.~1 into Eq.~\eqref{eq:dfdt}, we analytically calculate
the time derivative of the local magnetization,
whose maximum value is given by $|\frac{\partial}{\partial t}{\bm f}(\xx,t=0)|_{\rm max}\simeq 14.3 \hbar/(M\xi_{\rm knot}^2)$.
As the magnetization initially increases linearly as a function of time (see Fig. 3),
we define the lifetime of a knot as $\tau_{\rm knot} = [14.3 \hbar/(M\xi_{\rm knot}^2)]^{-1}$.
We also numerically calculate the dynamics of the knot shown in Fig.~1
by solving the time-dependent Gross-Pitaevskii equation; the result is shown in Fig.~3.
In Fig.~3, we plot the maximum polarization as a function of time.
The dashed line represents $|{\bm f}(\xx)|_{\rm max}=t/\tau_{\rm knot}$, which agrees well with the numerical results.
The insets of Fig.~3 show the distribution of $|{\bm f}(\xx)|$ on the $xz$ plane.
The ferromagnetic domain emerges because of the spin current caused by the spatial dependence of $\nn$,
and then expand outward.
In the case of a $^{23}$Na BEC, the lifetime for a knot with $\xi_{\rm knot}=10~\mu$m is $\tau_{\rm knot} =2.5$ ms.
The lifetime increases with increasing the size of knots.

\begin{figure}
\includegraphics[width=0.87\linewidth]{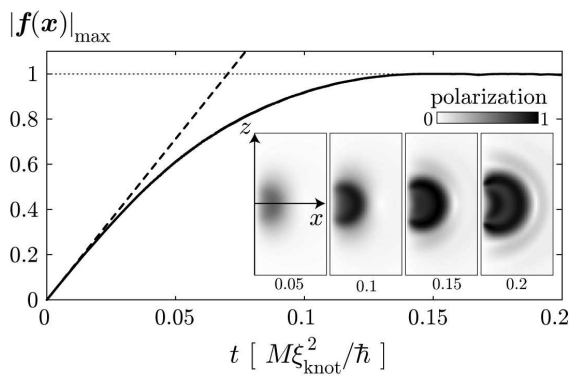}
\caption{
Time evolution of the maximum of the local polarization in a spherical trap with $R_{\rm TF}=4.3\xi_{\rm knot}$, starting from the
knot shown in Fig.~1.
The time is measured in units of $M\xi_{\rm knot}^2/\hbar$.
The dashed line shows $|{\bm f}(\xx)|_{\rm max}=t/\tau_{\rm knot}$ (see text), which agrees well with the numerical result.
The insets show $|{\bm f}(\xx)|$ in the $xz$ plane,
and the number below each panel shows the time elapsed in units of $M\xi_{\rm knot}^2/\hbar$.
The size of each panel is $1.9\xi_{\rm knot} \times 3.8\xi_{\rm knot}$.
}
\end{figure}

Finally, we point out that knots in a BEC may be used as an experimental signature of a magnetic monopole.
A magnetic monopole induces the magnetic field ${\bm B}=\hbar/(2 e r^2)\hat{r}$, which acts in a manner similar to the quadrupole field and creates knots.
Although $b(r)$ in this case diverges at $r=0$, it forms a knot-like structure on a large scale.
For instance, knots expand up to $10~\mu $m in the period $43~\mu$s of Larmor precession at $r=10~\mu$m, where $B=33$~mG.

In conclusion, we have shown that a spin-1 polar Bose-Einstein condensate can accommodate a knot,
which is also be shown to be created using a quadrupolar magnetic field.
Contrary to knot solitons known in other systems,
the knots in spinor BECs are immune from energetic instability against shrinkage, because the energy of the system is conserved;
however they are vulnerable to destruction caused by spin currents because of the $\nn$ texture.
The lifetime of a knot increases in proportion to the square of the size of knots
and is shown to be sufficiently long to be observed in a Stern-Gerlach experiment.

This work was supported by a Grant-in-Aid for Scientific Research (Grant No.\ 17071005)
and by a 21st Century COE program at Tokyo Tech ``Nanometer-Scale Quantum Physics'' from the
Ministry of Education, Culture, Sports, Science and Technology of Japan.


\end{document}